\documentclass[a4paper,fleqn,usenatbib]{mnras}

\usepackage{newtxtext,newtxmath}

\usepackage[T1]{fontenc}
\usepackage{ae,aecompl}

\usepackage{graphicx}	
\usepackage{amsmath}	
\usepackage{amssymb}	
\usepackage{url}
\usepackage{epstopdf}
\usepackage{float}
\usepackage{longtable}
\usepackage{caption}
\usepackage{subcaption}
\usepackage{mathtools}
\usepackage{tablefootnote}
\usepackage{booktabs}
\usepackage{makecell}
\usepackage{geometry}
\usepackage{lineno}

\newcommand{\xmm}{{\it XMM~\/}}
\newcommand{\xmmn}{{\it XMM-Newton~\/}}
\newcommand{\swift}{{\it Swift~\/}}

\def\Msun{\hbox{$\rm ~M_{\odot}$}}
\def\ergcms{{\rm ~erg~cm^{-2}~s^{-1}}}
\def\ergsec{{\rm ~erg~s^{-1}}}

\newcommand{\source}{FBQS~J1644+2619~}

\title[Multiwavelength properties of FBQS J1644+2619]{FBQS J1644+2619: multiwavelength properties and its place in the class of $\gamma$-ray emitting Narrow Line Seyfert 1s\thanks{Based on observations with XMM-Newton, an ESA Science Mission with instruments and contributions directly funded by ESA Member states and the USA (NASA)}}

\author[J. Larsson]{
J. Larsson,$^{1}$\thanks{E-mail: josla@kth.se} F. D'Ammando,$^{2, 3}$  S. Falocco,$^{1}$ M. Giroletti,$^{3}$ M. Orienti,$^{3}$ E. Piconcelli,$^{4}$
\newauthor  and S. Righini$^{3}$\\
$^{1}$ KTH, Department of Physics, and the Oskar Klein Centre, AlbaNova, SE-106 91 Stockholm, Sweden\\
$^{2}$ Dipartimento di Fisica e Astronomia, Universita` di Bologna, Via Gobetti 93/2, 40129 Bologna, Italy \\
$^{3}$ INAF, Istituto di Radio Astronomia di Bologna, Via P. Gobetti 101, 40129 Bologna, Italy\\
$^{4}$ Osservatorio Astronomico di Roma (INAF), via Frascati 33, 00040 Monte Porzio Catone (Roma), Italy \\
\\
}

\date{Accepted XXX. Received YYY; in original form ZZZ}

\pubyear{2018}

\begin{document}
\label{firstpage}
\pagerange{\pageref{firstpage}--\pageref{lastpage}}
\maketitle

\begin{abstract}
A small fraction of Narrow Line Seyfert 1s (NLSy1s) are observed to be $\gamma$-ray emitters. Understanding the properties of these sources is of interest since the majority of NLSy1s are very different from typical blazars. Here, we present a multi-frequency analysis of FBQS~J1644+2619, one of the most recently discovered $\gamma$-ray emitting NLSy1s. We analyse an $\sim 80$~ks \xmmn observation obtained in 2017, as well as quasi-simultaneous multi-wavelength observations covering the radio -- $\gamma$-ray range. The spectral energy distribution of the source is similar to the other $\gamma$-ray NLSy1s, confirming its blazar-like nature. The X-ray spectrum is characterised by a hard photon index ($\Gamma = 1.66$) above 2~keV and a soft excess at lower energies.The hard photon index provides clear evidence that inverse Compton emission from the jet dominates the spectrum, while the soft excess can be explained by a contribution from the underlying Seyfert emission.  This contribution can be fitted by reflection of emission from the base of the jet, as well as by Comptonisation in a warm, optically thick corona. We discuss our results in the context of the other $\gamma$-ray NLSy1s and note that the majority of them have similar X-ray spectra, with properties intermediate between blazars and radio-quiet NLSy1s.

\end{abstract}




\section{Introduction}
\label{introduction}

NLSy1s  are a peculiar subclass of Active Galactic Nuclei (AGN), thought to be characterised by relatively low-mass supermassive black holes ($\sim 10^6 - 10^7\ \Msun$) and high accretion rates \citep{Boroson2002,Collin2004,Zhou2006}. They are defined based on the properties of the optical spectra, where they have FWHM(H$\beta) \leq 2000\ {\rm km\ s^{-1}}$, O[III]/H$\beta < 3$ and strong Fe~II emission \citep{Osterbrock1985}. They also differ somewhat from regular Seyfert~1s in the X-ray band, where they have steep spectra, strong soft excess and rapid variability \citep{Boller1996,Leighly1999a,Leighly1999b}. Since the launch of the {\it Fermi} $\gamma$-ray mission, it has been established that a small fraction of these sources are $\gamma$-ray emitters. The $\gamma$-ray properties provide clear evidence for the existence of powerful jets close to the line of sight. To date, nine $\gamma$-ray NLSy1s have been detected at high significance: PMN~J0948+0022, 1H~0323+342, PKS~1502+036, PKS 2004-447, SBS~0846+513, FBQS~J1644+2619, SDSS~J122222.55+041315.7 (J1222+0413 from hereon), B3~1441+476  and NVSS~J124634+023808 \citep{Abdo2009a,Abdo2009b, D'Ammando2012,D'Ammando2015,Yao2015b,D'Ammando2016}. 

With spiral host galaxies, low black hole masses and large accretion rates, NLSy1s clearly stand out from 'typical' jetted AGN, which are commonly found in elliptical galaxies harbouring massive black holes (e.g., \citealt{Sikora2007}). It has also been found that a lower fraction of NLSy1s are radio loud (RL, 7~per~cent, \citealt{Komossa2006}) compared to $\sim$10 -- 20~per~cent for broad-line AGN \citep{Kellermann1989,Jiang2007}. Determining the physical properties of the $\gamma$-ray NLSy1s is thus interesting in terms of understanding the range of conditions under which jets can form. Despite having jets aligned close to the line of sight, X-ray observations of some of these sources have revealed that the accretion disc + corona likely contribute to the X-ray spectra \citep{D'Ammando2014,Paliya2014,Yao2015a}. Increasing the number of $\gamma$-ray NLSy1s with good-quality X-ray spectra is important for better understanding the connection between the jet and accretion flow in these sources. In this paper we analyse a long \xmmn observation of \source obtained in 2017. We also present quasi-simultaneous multiwavelength observations covering radio to $\gamma$-rays obtained with the Medicina radio telescope, the Rapid Eye Mount (REM) telescope, \swift and {\it Fermi} Large Area Telescope (LAT).

 \source ($z=0.145$, \citealt{Bade1995}) was first detected in $\gamma$-rays in 2015 by the {\it Fermi} LAT \citep{D'Ammando2015}. It has an average $\gamma$-ray luminosity of $1.6 \times 10^{44}\ \ergsec$ and an average photon index $\Gamma_{\gamma}=2.5$. The $\gamma$-ray light curve shows episodes of flaring activity, with the strongest one reaching a flux nine times higher than the average  \citep{D'Ammando2015}. These properties are similar to the other $\gamma$-ray NLSy1s, as well as Flat Spectrum Radio Quasars (FSRQs).  The radio properties of the source also confirm its blazar-like nature. Similar to the other $\gamma$-ray emitting NLSy1s, it has a high radio loudness\footnote{Defined as the ratio of flux densities at 5~GHz and the $B$-band} ($\rm log\ R=2.39$) and a flat radio spectrum \citep{Doi2012}. On the pc scale it has a one-sided core-jet structure with a very high core dominance, while the kpc scale shows a two-sided structure reminiscent of a Fanaroff-Riley type II radio galaxy \citep{Doi2011,Doi2012}. From the core dominance \cite{Doi2012} estimate a jet speed of $\beta = 0.983$ and a viewing angle $\theta < 5^{\circ}$.  

Previous observations of \source in the X-ray range have all had short exposures. A 3~ks {\it Chandra}-ACIS observation was obtained in 2003 and a \swift snapshot observation (1.3~ks) was performed in 2011. The 0.3--5~keV {\it Chandra} spectrum was equally well fitted with a power law with $\Gamma=2.2$ as with a soft-excess component  together with a flatter power law with $\Gamma=1.8$ \citep{Yuan2008}. The \swift X-Ray Telescope (XRT) 0.3--10~keV spectrum from 2011 is well fitted by a power law with $\Gamma=2.0 \pm 0.3$ \citep{D'Ammando2015}. 

This paper is organized as follows. We present the observations in Section \ref{data} and then present the results regarding the X-ray and multi-wavelength properties in Sections \ref{xray} and \ref{multiw}, respectively. We finally discuss the results and present our conclusions in Sections \ref{discussion} and \ref{conclusions}.  Throughout the paper we assume a flat cosmology with $H_0 = 70\ {\rm km\ s^{-1}\ Mpc^{-1}}$ and $\Omega_{\Lambda} = 0.73$. Uncertainties on fit parameters from {\sc xspec} are quoted at $90\%$ significance for one interesting parameter ($\Delta \chi^2$ =2.7). All other uncertainties are one sigma.

\section{Multiwavelength observations and data reduction}
\label{data}

All multiwavelength observations and data reduction procedures are described below, starting from the lowest frequencies. In the case of the $\gamma$-ray observations by {\it Fermi LAT} we only obtain an upper limit. The analysis performed to determine the upper limit is described in Section \ref{FermiData}.

\subsection{Medicina radio telescope observations}
\label{medicina}

We observed \source with the Medicina 32-m radio telescope on 2017 March 4 for a total net time on source of $\sim~6$ min. The observations were carried out in full polarization at a central frequency of 24.1 GHz, with two sub-bands (left and right circular polarization), each of 1.2~GHz width. Cross scans in declination and right ascension were also executed. We carried out the usual calibration procedures, setting the amplitude scale on 3C123, 3C286, and NGC7027, and compensating for the sky opacity determined through an off-source full scan of the sky from high to low elevation. This is similar to the procedures described in e.g., \cite{Egron2017}.
 
For each combination of sub-band and scan direction, we combined all good-quality data to improve the signal-to-noise ratio. The source was marginally detected in each of these individual combinations of sub-bands and scan directions. A weighted average of the measurements provides a $\sim 4 \sigma$ detection and a final estimate of the 24~GHz flux density of ($110\pm30$)~mJy.

\subsection{REM observations}
\label{remobs}

\begin{table*}
\caption{Log and optical magnitudes from the  REM observations of FBQS J1644$+$2619.}
\begin{center}
\begin{tabular}{ccccc}
\hline
\multicolumn{1}{c}{Date (UT)} &
\multicolumn{1}{c}{MJD} &
\multicolumn{1}{c}{$V$} &
\multicolumn{1}{c}{$R$} &
\multicolumn{1}{c}{$I$} \\
\multicolumn{1}{c}{} &
\multicolumn{1}{c}{} &
\multicolumn{1}{c}{(mag)} &
\multicolumn{1}{c}{(mag)} &
\multicolumn{1}{c}{(mag)} \\
\hline
2017-Mar-03 & 57815 & $17.317 \pm 0.042$ & $16.804 \pm 0.035$ & $16.073 \pm 0.079$ \\
2017-Mar-04 & 57816 & $17.184 \pm 0.037$ & $16.754 \pm 0.041$ & $16.179 \pm 0.087$ \\
2017-Mar-05 & 57817 & $17.329 \pm 0.036$ & $16.827 \pm 0.041$ & $16.117 \pm 0.074$ \\
2017-Mar-07 & 57819 & $17.472 \pm 0.032$ & $16.974 \pm 0.043$ & $16.271 \pm 0.065$ \\
2017-Mar-08 & 57820 & $17.625 \pm 0.064$ & $17.106 \pm 0.042$ & $16.363 \pm 0.085$ \\
2017-Mar-09 & 57821 & $17.654 \pm 0.056$ & $17.103 \pm 0.044$ & $16.298 \pm 0.091$ \\
2017-Mar-10 & 57822 & $17.237 \pm 0.052$ & $16.743 \pm 0.040$ & $16.049 \pm 0.083$ \\
2017-Mar-30 & 57842 & $17.495 \pm 0.046$ & $17.067 \pm 0.055$ & $16.497 \pm 0.078$ \\
\hline
\end{tabular}
\end{center}
\label{REM}
\end{table*}

\source was observed by REM during 2017 March 3--30 as part of a long-term project for AOT34 (PI: D'Ammando). REM \citep{Zerbi2001, Covino2004} is a robotic telescope located at the ESO Cerro La Silla observatory (Chile). It has a Ritchey-Chretien configuration with a 60-cm f/2.2 primary and an overall f/8 focal ratio in a fast moving alt-azimuth mount, providing two stable Nasmyth focal stations. At one of the two foci, the telescope simultaneously feeds, by means of a dichroic, two cameras: REMIR \citep{Conconi2004}  for the near-infrared (NIR) and ROSS2 \citep{Tosti2004} for the optical. The cameras both have a field of view of 10 $\times$ 10 arcmin and imaging capabilities with the usual NIR (z, $J$, $H$, and $K$) and Johnson-Cousins $VRI$ filters. The REM software system  \citep{Covino2004} is able to manage complex observational strategies in a fully autonomous way. 

All raw optical/NIR frames obtained with REM were reduced following standard procedures. Instrumental magnitudes were obtained via aperture photometry and absolute calibration was performed by means of secondary standard stars in the field reported by APASS\footnote{https://www.aavso.org/apass/} and by 2MASS\footnote{http://www.ipac.caltech.edu/2mass/} for the optical and NIR filters, respectively. We averaged the values obtained during the same observing night. The observed optical magnitudes are reported in Table \ref{REM}. The NIR detections turned out to be of low significance and are therefore not reported. The flux densities, corrected for extinction using the $E(B-V)$ value of 0.073 from \citet{schlafly11} and the extinction laws from \citet{cardelli89}, are presented in Section \ref{multiw}.

\subsection{\textbf{ \emph{ Swift} observations}}
\label{SwiftData}

\begin{table*}
\caption{Results of the {\em Swift}-UVOT and \xmmn Optical Monitor (OM) data for FBQS J1644$+$2619. The OM observation is marked by a *. }
\begin{center}
\begin{tabular}{cccccccc}
\hline
\multicolumn{1}{c}{Date (UT)} &
\multicolumn{1}{c}{MJD} &
\multicolumn{1}{c}{$v$} &
\multicolumn{1}{c}{$b$} &
\multicolumn{1}{c}{$u$} &
\multicolumn{1}{c}{$w1$} &
\multicolumn{1}{c}{$m2$} &
\multicolumn{1}{c}{$w2$} \\
\multicolumn{1}{c}{} &
\multicolumn{1}{c}{} &
\multicolumn{1}{c}{(mag)} &
\multicolumn{1}{c}{(mag)} &
\multicolumn{1}{c}{(mag)} &
\multicolumn{1}{c}{(mag)} &
\multicolumn{1}{c}{(mag)} &
\multicolumn{1}{c}{(mag)} \\
\hline
2015-Apr-09 & 57121 & 17.46$\pm$0.22 & 18.07$\pm$0.18 & 17.37$\pm$0.16 & 17.27$\pm$0.17 & 17.62$\pm$0.19 & 17.28$\pm$0.09 \\
2015-May-05 & 57147 & 17.84$\pm$0.28 & 18.21$\pm$0.19 & 17.25$\pm$0.15 & 17.45$\pm$0.11 & 17.53$\pm$0.19 & 17.48$\pm$0.14 \\
2015-Jun-05 & 57178 & - & - & - & 17.45$\pm$0.11 & 17.54$\pm$0.12 & - \\
2015-Jul-05 & 57208 & - & - & 17.16$\pm$0.07 & - & - & -\\
2015-Aug-05 & 57239 & 17.55$\pm$0.21 & 17.78$\pm$0.15 & 16.85$\pm$0.13 & 16.97$\pm$0.11 & 17.03$\pm$0.14 & 16.90$\pm$0.11 \\
2015-Sep-05 & 57270 & 17.74$\pm$0.20 & 18.09$\pm$0.18 & 17.20$\pm$0.13 & 17.08$\pm$0.12 & 17.07$\pm$0.12 & 17.25$\pm$0.10 \\
2017-Feb-27 & 57811 & 17.34$\pm$0.14 & 17.96$\pm$0.11 & 16.87$\pm$0.08 & 16.85$\pm$0.10 & 16.99$\pm$0.11 & 17.01$\pm$0.09 \\
2017-Mar-03* & 57815 & 17.08$\pm$0.02 & 18.04$\pm$0.02 & 16.96$\pm$0.01 & 16.70$\pm$0.02 & 17.08$\pm$0.05 & 17.20$\pm$0.08 \\
2017-Mar-07 & 57819 & 17.42$\pm$0.13 & 17.96$\pm$0.11 & 17.02$\pm$0.09 & 17.14$\pm$0.11 & 17.34$\pm$0.12 & 17.17$\pm$0.09 \\
\hline
\end{tabular}
\end{center}
\label{UVOT}
\end{table*}

The {\em Swift} satellite \citep{gehrels04} carried out eight short (1.3--2.9~ks) observations of
FBQS J1644$+$2619 between 2015 April 9 and 2017 March 7, as listed in Table~\ref{UVOT}. The observations were
performed with all three instruments on board: the XRT \citep[][0.2--10.0~keV]{burrows05}, the Ultraviolet/Optical Telescope \citep[UVOT;][170--600~nm]{roming05} and the Burst Alert Telescope \citep[BAT;][15--150~keV]{barthelmy05}. The hard X-ray flux of the source turned out to be below the sensitivity of the BAT
 for these short exposures and the data from this instrument will therefore not be used.
Moreover, the source was not included in the {\em Swift} BAT 70-month hard
X-ray catalogue \citep{baumgartner13}. 

The XRT data were processed with standard procedures (\texttt{xrtpipeline
  v0.13.3}), filtering, and screening criteria by using the \texttt{HEAsoft}
package (v6.20). The data were collected in photon counting mode in all the
observations. The source count rate was low ($<$ 0.5 counts s$^{-1}$); thus
pile-up correction was not required. Source events were extracted from a circular region with a radius of 20 pixels (1 pixel $\sim$ 2.36 arcsec), while background events were extracted from a circular region with a radius of 50 pixels far away from the source region. Ancillary response files were generated with \texttt{xrtmkarf}, and account for different extraction regions, vignetting and point spread function
corrections. We used the spectral redistribution matrices v014 in the
Calibration data base maintained by \texttt{HEASARC}. Considering the low
number of photons collected ($<$ 200 counts per observation) the spectra were rebinned with a
minimum of 1 count per bin and Cash statistics \citep{cash79} was used for the spectral analysis. 

During the {\em Swift} pointings, the UVOT instrument observed FBQS J1644$+$2619
in all its optical ($v$, $b$ and $u$) and UV ($w1$, $m2$ and $w2$) photometric
bands \citep{poole08,breeveld10}. We analysed the data using the
\texttt{uvotsource} task included in the \texttt{HEAsoft} package (v6.20). Source
counts were extracted from a circular region of 5 arcsec radius centred on
the source, while background counts were derived from a circular region of
10 arcsec radius in a nearby source-free region. The observed magnitudes are
reported in Table~\ref{UVOT}. The UVOT flux densities were
corrected for extinction as described for the REM observations in Section \ref{remobs}.

\subsection{XMM-Newton observation}

\source was observed by \xmmn  \citep{Jansen2001} between 2017 March 3--4 for a total duration of 82~ks (OBS ID: 0783230101, PI: J. Larsson). All three EPIC cameras (pn, MOS1 and MOS2) were operated in Large Window mode. The data were reduced using the \xmmn Science Analysis System ({\small SAS v16.0.0}) following standard procedures. Strong background flaring was present intermittently throughout the observation. These time intervals were filtered out following standard procedures\footnote{https://www.cosmos.esa.int/web/xmm-newton/sas-thread-epic-filterbackground} using the high-energy light curves with cuts of 0.4 and 0.35 counts~$\rm{s^{-1}}$ for the pn and MOS, respectively. Varying the cuts in the range 0.3--0.6 counts~$\rm{s^{-1}}$ did not significantly affect the resulting spectra. The total good exposure times after the filtering are 47, 59 and 62~ks for the pn, MOS1 and MOS2, respectively. Source and background spectra were extracted from circular regions of radius 34 arcsec for all three detectors. All spectra were binned to contain at least 25 counts per bin and not to oversample the intrinsic energy resolution by more than a factor of three. The resulting 0.3--10~keV spectra contain approximately $42000$, $12000$ and $15000$ net source counts for the pn, MOS1 and MOS2, respectively. The background level is $\sim 1.5$~per~cent over the full energy interval for all detectors and $9-17$~per~cent above 6~keV (where the lowest background is for the pn and the highest for MOS1).

The data from the two Reflection Grating Spectrometers (RGS) were reduced using \texttt{rgsproc}. Both detectors have 57~ks of good exposure time after removing the time intervals with high background. Merging the first order spectra of RGS1 and RGS2 results in a spectrum with $\sim 3000$ net source counts over 0.3--2~keV. The background level is $\sim 50$~per~cent over the whole energy interval and $85$~per cent below 0.5~keV. No lines were detected and the spectra do not have sufficient quality to discriminate between the different models presented in section \ref{xmmtav}. The RGS data will therefore not be discussed further.

The OM observed the source in all six filters in imaging mode together with a fast readout window. The total exposure times of the imaging observations are: 8800~s ($v$), 8000~s ($b$), 8200~s ($u$), 17600~s ($w1$), 17600~s ($m2$) and 19800~s ($w2$).  The data were processed using the {\small SAS} tasks \texttt{omichain} and \texttt{omfchain}. The observed average magnitudes for the imaging mode are reported in Table \ref{UVOT} together with the \swift UVOT magnitudes. The flux densities were corrected for extinction as described for REM in Section \ref{remobs}.  The size of the time bins for the fast mode was set to $100$~s for all filters.

\subsection{{\em Fermi}-LAT observations}
\label{FermiData}

\source is regularly observed in $\gamma$ rays as part of the ongoing sky survey by the {\em Fermi}-LAT.  The {\em Fermi}-LAT  is a pair-conversion telescope operating from 20~MeV to $>$ 300~GeV \citep{atwood09}. The LAT data used in this paper were collected from 2017 February 17 to March 18. During this time, the LAT instrument
operated almost entirely in survey mode. The Pass 8 data,\footnote{https://fermi.gsfc.nasa.gov/ssc/data/analysis/documentation/\\Pass8\_usage.html}
 based on a complete and improved revision of the entire LAT event-level
analysis, were used. The analysis was performed with the \texttt{ScienceTools} software package version v10r0p5. We used only events belonging
to the `Source' class (\texttt{evclass=128}), including front and back converting events (\texttt{evtype=3}). Events were selected within a maximum zenith angle of $90^{\circ}$ to
reduce contamination from the Earth-limb $\gamma$ rays, which are produced by cosmic rays interacting with the upper atmosphere. 
The spectral analysis was performed with the instrument response functions \texttt{P8R2\_SOURCE\_V6}, using a binned maximum-likelihood method implemented in the Science tool \texttt{gtlike}. Isotropic (`iso\_source\_v06.txt') and Galactic diffuse emission (`gll\_iem\_v06.fit') components were used to model the background \citep{acero16}\footnote{http://fermi.gsfc.nasa.gov/ssc/data/access/lat/\\BackgroundModels.html}. The normalisations of both components were allowed to vary freely during the spectral fitting.

We analysed a region of interest of $30^{\circ}$ radius centred at the location of FBQS J1644$+$2619. We evaluated the significance of the $\gamma$-ray signal from the source by means of a maximum-likelihood test statistic (TS) defined as TS = 2$\times$(log$L_1$ - log$L_0$), where $L$ is the likelihood of the data given the model with ($L_1$) or without ($L_0$) a point source at the position of FBQS J1644$+$2619 \citep[e.g.,][]{mattox96}. The source model used in \texttt{gtlike} includes all the point sources from the 3FGL catalogue that fall within $40^{\circ}$ of FBQS J1644$+$2619. The spectra of these sources were parametrized by a power-law, a log-parabola, or a super exponential cut-off, as in the 3FGL catalogue. We also included new candidates within $7^{\circ}$ of FBQS J1644$+$2619 from a preliminary source list using 7 years of Pass 8 data.

A first maximum likelihood analysis was performed over the whole period to remove sources with TS $< 10$ from the model. A second maximum likelihood analysis was then performed on the updated source model. In the fitting procedure, the normalisation factors and the spectral parameters of the sources lying within 10$^{\circ}$ of FBQS J1644$+$2619 were left as free parameters. For the sources located between 10$^{\circ}$ and 40$^{\circ}$ from our target, we kept the normalisation and the spectral shape parameters fixed at the values from the 3FGL catalogue.

Integrating over 2017 February 17 -- March 18, the fit with a power-law model ($dN/dE\propto$ $(E/E_{0})^{-\Gamma_{\gamma}}$) results in TS = 1 in the 0.1--300~GeV energy range. The 2$\sigma$ upper limit is 1.44$\times$10$^{-8}$ ph cm$^{-2}$ s$^{-1}$, assuming a photon index of $\Gamma=2.5$.

\section{X-ray properties}
\label{xray}

\subsection{XMM-Newton observation}

\begin{figure}
\resizebox{80mm}{!}{ \includegraphics{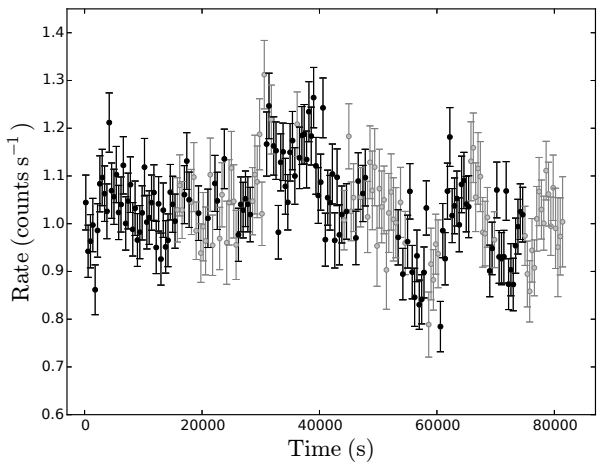}}
\caption{\xmmn EPIC~pn light curve over 0.3--10~keV with 400~s bins. The time intervals in grey are affected by strong background flares.}
 \label{lcurve}
 \end{figure}

The light curve of the observation is shown in Fig.~\ref{lcurve}. The grey data points show the time intervals that are affected by background flares and hence excluded from the spectral analysis. The light curve shows only moderate variability. The fractional root-mean-square variability\footnote{Defined as $F_{\rm var} = \sqrt{\frac{S^2 - \overline{\sigma_{\rm err}^2}}{\bar{x}^2} } $, where $S$ is the variance, $\sigma_{\rm err}$ is the mean error and $\bar{x}$ is the mean count rate. See \cite{Vaughan2003} for details.} calculated below 2~keV (where the background is low even during the flaring intervals)  is $0.087\pm 0.01$. Below we first analyse the time-averaged spectrum and then address the spectral variability. 

\subsubsection{Time-averaged spectrum}
\label{xmmtav}

\begin{figure}
\resizebox{80mm}{!}{ \includegraphics{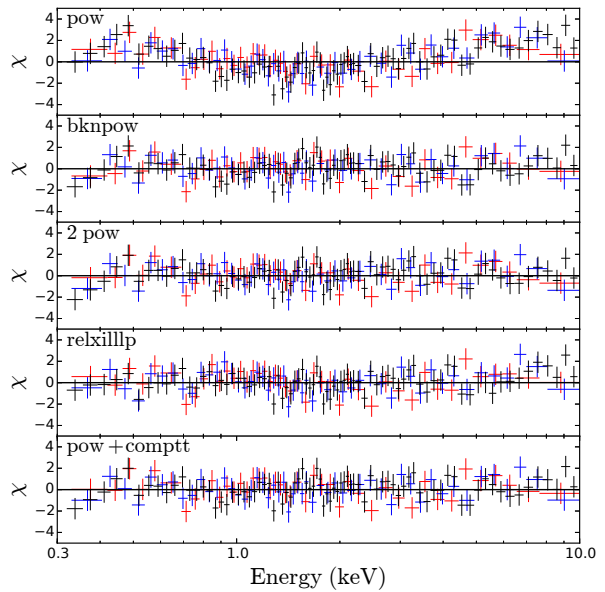}}
\caption{Residuals between the data and the power law, broken power law, double power law, {\sc relxilllp} and power law+{\sc comptt} models. The best-fitting parameters are provided in Table~\ref{xmmfits}. Data points from pn, MOS1 and MOS2 are shown in black, red and blue, respectively.}
 \label{residuals}
 \end{figure}

The spectral analysis was performed using {\sc xspec}~v12.8.2. We fitted the pn, MOS1 and MOS2 spectra simultaneously over the 0.3--10~keV energy range. All fit parameters were tied between the spectra except for a cross-normalisation constant. The value of the latter was always in the range 1.03-1.05. All fits include Galactic absorption fixed at $5.20\times 10^{20}\ {\rm cm^{-2}}$ \citep{Kalberla2005} using the {\sc tbabs} model. The fit results are presented in Table \ref{xmmfits} and the residuals of the different models are shown in Fig.~\ref{residuals}. 

As these results show, the spectrum is clearly inconsistent with a pure power law, while a broken power law provides a good fit. In this model the spectrum breaks from a softer to a harder slope ($\Gamma_1=1.90\pm 0.02$ and $\Gamma_2=1.66^{+0.03}_{-0.04}$) at an energy $E_{\rm break} =1.9^{+0.3}_{-0.2}$~keV.  Adding intrinsic absorption to the model does not improve the quality of the fit. As expected from these results, the 2--10~keV spectrum is well fitted by a single power law with $\Gamma=1.66\pm 0.04$ ($\chi^2$/d.o.f = 244/235). The 0.3--10~keV (2--10~keV) unabsorbed flux obtained from the broken power-law model is $ 3.34 \pm 0.04 \times 10^{-12} \ergcms$  ($1.86 \pm 0.03 \times 10^{-12} \ergcms$), which corresponds to a luminosity of $2.0 \times 10^{44} \ergsec$ ($1.1 \times 10^{44}\ergsec$).

\begin{table}
\begin{center}
\begin{tabular}{lll}
\hline \\ [-4pt]
Model& Parameter & Value
 \\[4pt]  
\hline\\[-6pt]

Power law & $\Gamma$                                 & $1.82\pm 0.01$  \\
                    & Norm                                                 &  $5.07\pm 0.04 \times 10^{-4}$     \\
                    &  $\chi^2/\rm{d.o.f.}$                   &    478/355  \\[4pt]  
\hline 
Broken power law & $\Gamma_1$               & $1.90\pm 0.02$   \\
& ${E_{\rm break}}$ (keV)                                & $1.9^{+0.3}_{-0.2}$   \\
& $\Gamma_2$                                                 & $1.66^{+0.03}_{-0.04}$  \\
& Norm                                                                     & $5.00\pm 0.05 \times 10^{-4}$   \\
& $\chi^2/\rm{d.o.f.}$                                        &  349/353\\[4pt]  
\hline
Power law & $\Gamma_1$               & $2.01^{+0.14}_{-0.07}$   \\
+ power law  & Norm$_1$             & $4.4^{+0.4}_{-1.0} \times 10^{-4}$   \\
& $\Gamma_2$                                         & $1.0^{+0.3}_{-0.4}$  \\
& Norm$_2$                                  & $6^{+10}_{-4} \times 10^{-5}$   \\
& $\chi^2/\rm{d.o.f.}$                                        &  348/353\\[4pt]  
\hline
relxilllp   &		$h$ $(r_{\rm{g}})$	&    $45^{*}_{-34}$ \\
&	$a$	&  0.998$^{f}$ \\
 &  incl    ($^\circ$)	& $5^{f}$    \\
&  $R_{\rm{in}} (r_{\rm g})  $ & $1.4^{+40.8}_{*}$ \\
&  $R_{\rm{out}}\ (r_{\rm{g}})$	& $400^{f}$\\			        
& $\Gamma$   &    $1.78\pm 0.01$ \\
&  log $\xi\ (\ergcms)$	&	 $1.6^{+0.3}_{-0.2}$ \\
&  ${A_{\rm Fe}}$  &  $1^{f}$ \\
& ${E_{\rm cut}}$ (keV)  & $300^{f}$ \\	
& R & $0.88^{*}_{-0.15}$	\\	        
& Norm  & $1.2^{+3.8}_{-0.2} \times 10^{-5}$     \\
&  $\chi^2/\rm{d.o.f.}$                   &    355/351  \\[4pt]  
   \hline
Power law  & $\Gamma$                                 & $1.64^{+0.05}_{-0.08}$  \\
 + compTT                   & PL Norm                                                 &  $4.18^{+0.03}_{-0.05} \times 10^{-4}$     \\
 	& ${kT}_{\rm 0}$ (eV) & $26^f$ \\
  	& ${kT}_{\rm e}$ (keV) & $0.30^{+0.21}_{-0.11}$ \\
	& $\tau$ & $16^{+11}_{-4}$ \\
	& Norm  &  $1.4^{+0.3}_{-0.5} \times 10^{-2}$ \\
 
                    &  $\chi^2/\rm{d.o.f.}$                   &    346/352  \\[4pt]  
   
\hline \\  
   
\end{tabular}
\caption{\label{xmmfits}\small{Summary of fits to the \xmmn spectra.  All
    fits also include absorption fixed at the Galactic value. Superscript $f$ indicates that a parameter was kept fixed.  A * indicates that the confidence interval reached the hard boundary of a parameter.}}
\end{center}
\end{table}

The photon index above 2~keV is significantly harder than in radio-quiet NLSy1s (e.g. \citealt{Grupe2010,Foschini2015}) and instead similar to values measured in radio-loud AGN (e.g., \citealt{Piconcelli2005}). This is a clear indication that emission from the jet dominates the spectrum, as expected given the other blazar-like properties of the source. The photon index below 2~keV is also hard compared to typical radio-quiet sources. However, the fact that the spectrum softens at low energies indicates that an emission component in addition to the jet may be present. A simple such two-component model consists of two power laws, originating from a standard accretion disc corona and a jet, respectively. The photon indices obtained when fitting this model are $\Gamma_1=2.01^{+0.14}_{-0.07}$ and $\Gamma_2=1.0^{+0.3}_{-0.4}$. While the former value is typical of AGN coronae, the second photon index is extremely hard even for a blazar. It is thus motivated to explore more complex models.

From a physical perspective, part of the radiation emitted from the base of the jet will irradiate the accretion disc, giving rise to a 'reflection spectrum' due to a combination of Compton scattering and fluorescence, which is relativistically blurred if arising from the inner disc. The clearest signature of reflection in AGN is the Fe K$\alpha$ line at 6.4~keV, which is not detected in our observation. In particular, we obtain an upper limit on the equivalent width of $49$~eV ($90\%$ confidence) for a narrow ($\sigma = 10$~eV) line at 6.4~keV. Another signature of reflection in the 0.3--10~keV energy range is strong emission below $\sim 2$~keV due to the combined emission from a large number of relativistically broadened lines. This emission may make up part or all of the so-called soft excess commonly observed in AGN (e.g, \citealt{Crummy2006}).

In order to explore if the soft excess in \source can be explained by a contribution from reflection (while the Fe line may be undetected due to limited statistics and 'unfavourable' physical conditions of the disc/jet system) we use the {\sc relxilllp} model \citep{Dauser2013,Garcia2014}, which calculates the reflection spectrum from a point source located on the rotational axis above the black hole. In our physical scenario, this source would correspond to the base of the jet. As the emission from the jet is expected to be beamed away from the disc, the flux that reaches the disc will be reduced compared to the case of a stationary point source. Since our tests also showed that the height of the primary source is large, we constrained the reflection fraction ($R$) to be $<1$ in the fits, where $R$ is defined as the ratio of primary-source intensity illuminating the disc to the primary-source intensity that reaches the observer \citep{Dauser2016}. The model also includes the power-law emission from the point source that directly reaches the observer. The effects of approximating the jet by a point source are discussed in Section \ref{xdiscussion}. 

The current observation in the limited band pass of \xmmn cannot constrain all the free parameters of this model. However, there are strong constraints on the inclination from radio observations ($i<5^{\circ}$, \citealt{Doi2012}) and we fix the inclination at $i=5^{\circ}$ after tests showed that allowing it to vary to lower values did not significantly change the results. Additionally, we fix the outer radius of the disc ($r_{\rm out}$), the iron abundance ($A_{\rm Fe}$) and the cutoff energy of the power-law ($E_{\rm cut}$) at their default values (see Table~\ref{xmmfits}). We note that the latter value is inappropriate given that the emission is from a jet, but that this choice does not affect the results since the cutoff is well outside the observed energy range. The value of $r_{\rm out}$  also has a very small effect on the other best-fitting parameters. When  $A_{\rm Fe}$ is allowed to vary it reaches the lowest allowed value of 0.5 and the fit improves by $\Delta \chi^2 = 5$, while the other best-fitting parameters do not change significantly. It is kept fixed because of its relatively small impact on the fit and because an increase in the number of free parameters causes problems with the convergence of the error calculations. Finally, we fix the black hole spin at the maximal value of $a=0.998$ after tests showed that we were unable to constrain this parameter. In particular, we note that we are still unable to constrain the spin if we fix the inner edge of the disc ($R_{\rm in}$) at the Innermost Stable Circular Orbit (ISCO). Assuming instead a non-spinning black hole ($a=0$) only has a small impact on the results, the most important of which is that the upper limit on $R_{\rm in}$ is about $10\ r_{\rm g}$ higher. After fixing these parameters, there are six free parameters of the model, as summarized in Table~\ref{xmmfits}.

\begin{figure}
\resizebox{80mm}{!}{\includegraphics{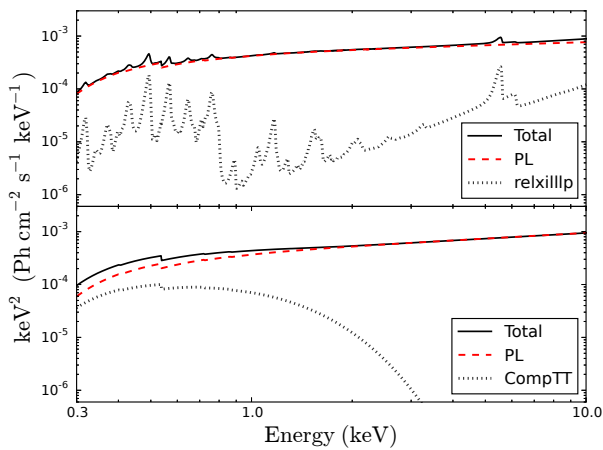}}
\caption{Best-fitting models to the \xmmn spectra for the cases where reflection (top panel) and Comptonisation (bottom panel) contribute to the spectra. The solid, black line is the total model, the red, dashed line is the power-law emission from the jet, while the dotted, grey line shows the contribution from reflection or Comptonisation. Both models also include Galactic absorption.}
\label{models}
\end{figure}

This model provides a good fit of comparable quality as the broken power-law and double power-law models. The ionisation parameter is low at log~$(\xi)\ =1.6^{+0.3}_{-0.2}\ \ergcms$, while the height of the primary source is large ($h> 11\  r_{\rm{g}}$) and the inner edge of the disc is constrained to be $R_{\rm{in}} <42\ R_{\rm r_g}$. In this model the reflection spectrum contributes only to about $7$~per~cent of the total flux in the 0.3--10~keV range. While this is sufficient to affect the curvature of the spectrum, it is low enough that the Fe line remains undetected. The reflection fraction is $R=0.88^{+0.12}_{-0.15}$, where the upper confidence interval reached the hard  boundary of 1.  When this constraint is removed, the best-fitting value is $R=0.88^{+0.17}_{-0.15}$, i.e. the upper limit is only slightly above 1. The contributions from the jet and reflection components are shown in the top panel of Fig~\ref{models}.

There are some clear limitations of this interpretation. First, the model assumes a stationary point source as the primary source, while the jet is an extended structure accelerated away from the disc. As discussed by \cite{Dauser2013}, the effect of neglecting this is rather limited when the source is far away from the disc (see further Section \ref{xdiscussion}). Second, while the fit shows that the observations are consistent with a contribution from reflection, the fact that we do not detect an Fe line means that other explanations cannot be ruled out. For example, we note that we also obtain a good fit to the spectra using a model comprising a power law and Comptonisation of the thermal emission from the disc by a warm, optically thick electron population (see bottom panels of Fig.~\ref{residuals} and Fig.~\ref{models}). This electron population may correspond to a heated upper layer of the accretion disc. We model the Comptonisation with the {\sc comptt} model \citep{Titarchuk94} using the disc geometry. The assumed geometry affects the resulting optical depth ($\tau$) but not the spectrum calculated by the model. We fix the temperature of the seed photons at $kT_{\rm 0} =26$~eV, corresponding to the temperature of a standard accretion disc for a $1.4 \times 10^7\  \Msun$ black hole \citep{Foschini2015} accreting at 0.2 $\times$ Eddington.\footnote{The accretion rate was estimated from $L_{\rm bol}=9\lambda L_{5100}$ with $\lambda L_{5100}$ estimated from the H$\beta$ luminosity as described in \cite{Zhou2006} in order to avoid contamination from the jet. The  H$\beta$ luminosity was taken from \cite{Foschini2015}.} The temperature ($ kT_{\rm e} $) and $\tau$ of the Comptonising electrons are left free to vary. We note that these parameters have a very weak dependence on the assumed value of  $kT_{\rm 0}$.  In particular, they do not change at all if we adopt a lower ${kT_{\rm 0}}$, corresponding to the higher black hole mass of $2 \times 10^8\  \Msun$ found by \cite{Calderone2013} and \cite{D'Ammando2017}, and correspondingly lower Eddington ratio (the black hole mass and accretion rate are further discussed in Section \ref{nature}). The only parameter that is affected by the exact value of $kT_{\rm 0}$ is the normalisation of the {\sc comptt} model. 
 
The Comptonisation and reflection scenarios differ in where they contribute most of the flux, as seen in Fig.~\ref{models}. The {\sc comptt} component only affects the low-energy part of the spectrum, making up $\sim$ 20~per~cent of the flux below 2~keV. The reflection spectrum instead affects the spectral curvature by predominantly contributing at the lowest and highest energies. These differences also affect the predicted hard X-ray flux. However, both models are consistent with the non-detection by BAT in the 14--195~keV energy range. Specifically, the sensitivity limit of the BAT 70-month survey  is $1 \times 10^{-11} \ergcms$ \citep{baumgartner13}, while the predicted flux in the BAT energy range is $6.4\times 10^{-12}$ and $7.4\times 10^{-12} \ergcms$ for the reflection and Comptonisation scenarios, respectively.\footnote{In these flux calculations we have assumed that none of the power-law components have a cut off.}  Even though the reflection model has a significant contribution from the Compton hump in the hard X-ray range, the total predicted flux is slightly lower due to the softer photon index (cf. Table \ref{xmmfits}). The reflection model predicts a higher flux than the Comptonisation model in the 10--40~keV range, near the peak of the Compton hump ($2.9 \times 10^{-12}$ compared to $2.7 \times 10^{-12} \ergcms$), but this is still below the corresponding BAT sensitivity of $4.3\times 10^{-12} \ergcms$ (\citealt{Ricci2015}, assuming a power-law spectrum with $\Gamma=1.8$). Finally, we note that the double power-law model predicts a 14--195~keV flux of $2 \times 10^{-11} \ergcms$,  which is above the sensitivity limit of the BAT survey.

\subsubsection{Time-resolved spectra}
\label{xmmvar}

\begin{figure}
\resizebox{80mm}{!}{ \includegraphics{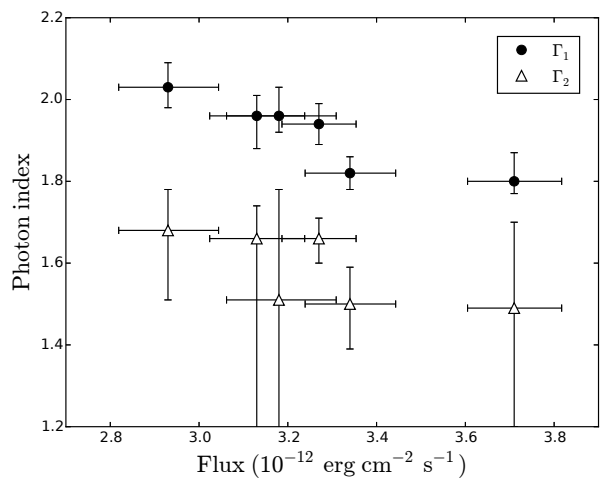}}
\caption{Results from fits to time-resolved spectra with a broken power law. The photon indices below and above the break are shown as filled circles and open triangles, respectively. The flux is the unabsorbed flux in the 0.3--10~keV range.}
 \label{tresfits}
 \end{figure}

The relatively low count rate together with the presence of background flares makes it impossible to perform a detailed  analysis of the variability in this source. 
In order to still place some constraints on the spectral variability, we fitted the broken power-law model to spectra extracted from 10~ks intervals of the light curve. Two of the resulting spectra have less than 4~ks of good exposure time (the intervals between 20-30~ks and 50-60~ks, cf.~Fig.~\ref{lcurve}). These spectra were excluded from the analysis since it was not possible to constrain all the parameters of the model with such short exposure times. The remaining six spectra have 3800--8300 counts in the pn, and about a factor 3 lower in each of the MOS detectors. The results from fitting these spectra show that only the photon index below the break ($\Gamma_1$) is significantly variable at the $3\sigma$ level. The break energy was poorly constrained in most spectra. Both photon indices are plotted as a function of the flux in Fig.~\ref{tresfits}, which shows some weak evidence of $\Gamma_1$ hardening with increasing flux (Pearson r and p-values of -0.91 an 0.012, respectively). 

\subsection{\textbf{ \emph{ Swift}} XRT observations}
\label{xrtresults}

\begin{table*}
\caption{Log and fitting results of {\em Swift}-XRT observations of FBQS J1644$+$2619 using a power-law model with $N_{\rm H}$ fixed to the Galactic value.}
\begin{center}
\begin{tabular}{ccccccc}
\hline
\multicolumn{1}{c}{Date (UT)} &
\multicolumn{1}{c}{MJD} &
\multicolumn{1}{c}{Net exposure time} &
\multicolumn{1}{c}{Photon index} &
\multicolumn{1}{c}{Flux 0.3--10 keV$^{\rm a}$} &
\multicolumn{1}{c}{Flux 2.0--10 keV}  &
\multicolumn{1}{c}{cstat/d.o.f} \\
\multicolumn{1}{c}{} &
\multicolumn{1}{c}{} &
\multicolumn{1}{c}{(s)} &
\multicolumn{1}{c}{($\Gamma_{\rm\,X}$)} &
\multicolumn{1}{c}{($\times$10$^{-12}$ erg cm$^{-2}$ s$^{-1}$)} &
\multicolumn{1}{c}{($\times$10$^{-13}$ erg cm$^{-2}$ s$^{-1}$)} &
\multicolumn{1}{c}{} \\
\hline
2015-Apr-09 & 57121 & 1651 & $1.85 \pm 0.38$ & $1.87^{+0.47}_{-0.41}$ & $9.8^{+4.2}_{-3.1}$ & 56/45 \\
2015-May-05 & 57147 & 1556 & $1.83 \pm 0.36$ & $1.66^{+0.46}_{-0.39}$ & $8.8^{+4.3}_{-2.3}$ & 34/42 \\
2015-Jun-05 & 57178 & 1314 & $1.78 \pm 0.43$ & $1.32^{+0.45}_{-0.37}$ & $7.3^{+2.7}_{-2.1}$ & 34/32 \\
2015-July-05 & 57208 & 1651 & $1.61 \pm 0.41$ & $1.63^{+0.53}_{-0.43}$ & $10.2^{+5.3}_{-3.5}$ & 39/37 \\
2015-Aug-05 & 57239 & 1988 & $1.84 \pm 0.23$ & $2.90^{+0.50}_{-0.45}$ & $15.3^{+4.1}_{-2.2}$ & 85/105 \\
2015-Sep-05 & 57270 & 1503 & $2.08 \pm 0.39$ & $1.25^{+0.36}_{-0.30}$ & $5.3^{+2.4}_{-1.6}$ & 29/39 \\
2017-Feb-27 & 57811 & 2909 & $1.62 \pm 0.22$ & $2.09^{+0.39}_{-0.35}$ & $12.9^{+2.3}_{-2.4}$ & 82/98 \\
2017-Mar-07 & 57819 & 2682 & $1.80 \pm 0.19$ & $2.94^{+0.45}_{-0.40}$ & $16.0^{+1.8}_{-2.7}$ & 99/127 \\
\hline
\end{tabular}
\end{center}
$^{\rm a}$Unabsorbed flux
\label{XRT}
\end{table*}

The results from fitting the XRT spectra with an absorbed power-law are presented in Table~\ref{XRT}. As for the \xmm spectra, the absorption was modelled with {\sc tbabs} and kept fixed at its Galactic value. The photon indices are in the range $\Gamma= 1.61 -2.08$, with a median value of 1.82. The latter is consistent with the result of fitting the \xmmn spectra with a power law. While a broken power-law was required to obtain an acceptable fit  to the \xmmn spectra (Section \ref{xmmtav}), the single power law provides good fits to the low count-rate XRT spectra. No spectral variability is detected, as expected from the relatively large error bars on $\Gamma$ (typically $20 \%$). We therefore co-added all the XRT spectra, resulting in a spectrum with a total exposure of 15.1~ks. A power-law fit to this spectrum results in $\Gamma = 1.78\pm 0.12$  with $\chi^2/\rm{d.o.f}=30/28$, consistent with the median of the fits to the individual spectra. Fitting a broken power law instead gives $\Gamma_1=1.90\pm 0.13$, $\Gamma_2=1.48^{+0.28}_{-0.23}$ and $E_{\rm break} =2.19^{+1.28}_{-0.62}$~keV with $\chi^2/\rm{d.o.f}=27/26$. These parameters are fully consistent with the \xmmn spectrum, although we note that it is not significantly preferred over the single power law ($p=0.29$ according to an F-test). 

\begin{figure}
\resizebox{85mm}{!}{ \includegraphics{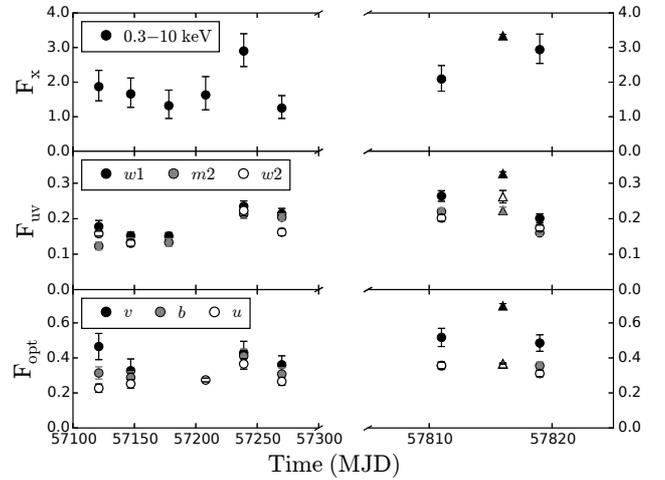}}
\caption{Flux evolution of \source during 2015 April 9 -- September 5 (first part of x-axis) and 2017 February 27 -- March 7 (second part of x-axis). Top panel: unabsorbed 0.3--10~keV flux in units of $10^{-12} \ergcms$, middle panel: extinction corrected UV fluxes in units of mJy (black: $w1$ band, grey: $m2$ band, white: $w2$ band), bottom panel: extinction corrected optical fluxes in units of mJy (black: $v$ band, grey: $b$ band, white: $u$ band). The \swift XRT and UVOT observations are shown as circles, while the \xmmn EPIC and OM observations are shown as triangles. Note that some of the error bars are smaller than the plot symbols.}
 \label{xrtlc}
 \end{figure}

The evolution of the 0.3--10~keV flux measured by XRT is shown in the top panel of Fig.~\ref{xrtlc}, together with the average flux of the \xmmn observation.  Note that the observations during 2015 were obtained at one-month intervals, while the observations in 2017 are separated by four days. The flux is clearly variable, with a ratio of 2.7 between the highest and lowest fluxes observed. The highest flux was observed during the \xmmn observation. At the end of the monitoring campaign in 2015, the flux approximately doubled and then decayed back to a similar level on a time-scale of a month. 

\section{Multi-wavelength properties}
\label{multiw}

\subsection{Optical and UV variability}

\begin{figure}
\resizebox{80mm}{!}{\includegraphics{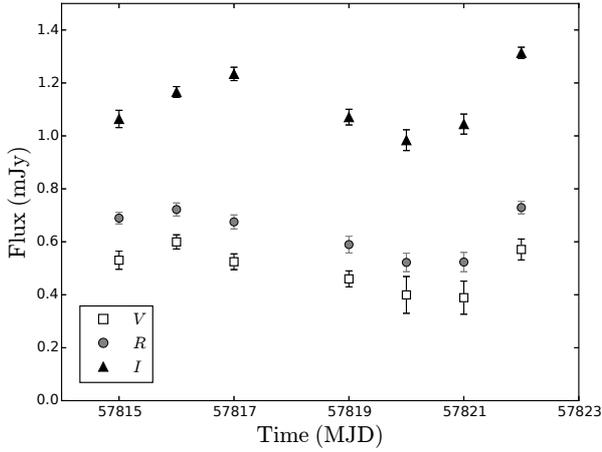}}
\caption{Optical light curve of \source from the monitoring with REM during 2017 Mar 3--10. Open squares, grey circles and black triangles show the extinction-corrected fluxes in the $V$, $R$ and $I$ bands, respectively.}
 \label{remlc}
 \end{figure}

Fig.~\ref{xrtlc} shows the time evolution of the UV ($w1,\ m2,\ w2$) and optical ($v,\ b,\ u$) fluxes probed by UVOT and the OM on a time-scale of months and days, together with the X-ray flux.  The UV and optical fluxes are variable, with the highest and lowest fluxes measured by UVOT differing by factors of 1.5--1.8 in the different bands. The \xmmn OM observation had the highest recorded fluxes in the $w1$, $w2$ and $v$ bands, where the flux in the latter was a factor $\sim 1.4$ higher than the fluxes measured by UVOT four days before and after.  Fig.~\ref{xrtlc} also shows an indication of a positive correlation between the X-ray and UV/optical fluxes. However, using Pearson's $r$-coefficient, the correlation with the X-ray flux is significant above the 2$\sigma$ level only for the $b$ band. All the optical and UV filters show marginal evidence of being correlated with each other, with significances in the $1-2\sigma$ range. The fast-mode OM data have low signal-to-noise and no significant variability was detected on short time-scales in these observations. 

The optical light curve from the monitoring with REM is shown in Fig.~\ref{remlc}. The monitoring started at the time of the \xmmn observation, i.e. 2017 March 3, and probes the flux on a one-day time-scale (except for one gap of two days) up to March 10 in the $V$, $R$ and $I$ bands. One additional observation taken 20 days later (not shown in the figure but included in Table~\ref{REM}) had a slightly lower flux level compared to the preceding observation. The fluxes in the three bands show a similar time evolution, with the $R$ and $V$ bands being correlated at $> 3 \sigma$. The maximum variability amplitudes are all around 1.5, i.e. similar to the UVOT results.

\subsection{Spectral energy distribution}

\begin{figure}
\resizebox{80mm}{!}{\includegraphics{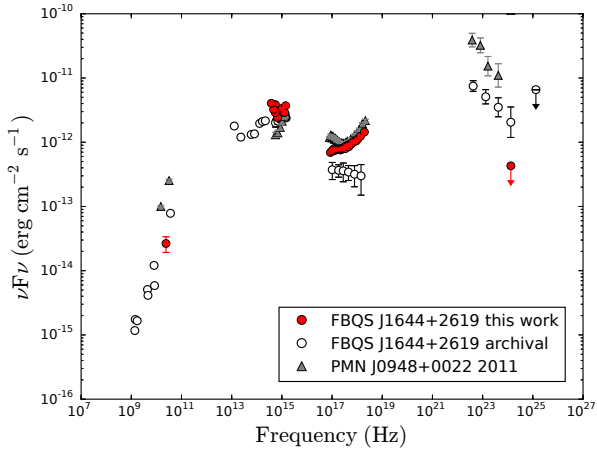}}
\caption{Spectral energy distribution of FBQS J1644$+$2619. Fluxes measured during the \xmmn observation are shown in red. These include observations from Medicina, REM,   \xmmn OM and EPIC pn, as well as an upper limit from {\it Fermi} LAT. Open symbols are previous observations collected from a variety of observatories and epochs (see text for details). For comparison, the SED of PMN J0948$+$0022 is also plotted (grey symbols, from \citealt{D'Ammando2015a}). The red and grey symbols have error bars that are too small to be seen in the optical--X-ray range. Uncertainties are not plotted for the the archival data in the radio--NIR range.}
\label{sed}
\end{figure}

The  spectral energy distribution (SED) of \source is shown in Fig.~\ref{sed}. The multi-wavelength data obtained quasi-simultaneously with the \xmmn observation are shown together with previous observations of the source. The latter include radio -- NIR fluxes reported by \cite{Foschini2015}, the {\it Swift} XRT and UVOT data from the lowest X-ray flux recorded in 2015 (section \ref{xrtresults}) as well as the largest $\gamma$-ray flux observed \citep[calculated using the time interval and spectral
parameters reported in][]{D'Ammando2015}. The upper limit on the $\gamma$-ray flux of $\rm{1.44 \times 10^{-8}\ ph\ cm^{-2}\ s^{-1}}$ obtained during the time of the \xmmn observation (see section \ref{FermiData}) is about a factor of 2.5 higher than the average flux of the source ($\rm{5.9 \times 10^{-9}\ ph\ cm^{-2}\ s^{-1}}$), but still significantly lower than the largest $\gamma$-ray flux recorded ($\rm{5.2 \times 10^{-8}\ ph\ cm^{-2}\ s^{-1}}$, \citealt{D'Ammando2015}). For comparison, we also show in Fig.~\ref{sed} the SED of the prototype $\gamma$-ray NLSy1 PMN~J0948+0022, as observed during an intermediate flux state in 2011 \citep{D'Ammando2015a}. While PMN~J0948+0022 has a larger luminosity than FBQS~J1644+2619, it also has a higher redshift, and the flux level of the two sources turn out to be similar in the radio -- X-ray range. The main difference between the SEDs is the much larger contribution at $\gamma$-ray energies for PMN~J0948+0022, revealing a higher Compton dominance in this source. While the average $\gamma$-ray and X-ray luminosities are similar in FBQS~J1644+2619, the luminosities differ by almost two orders of magnitude in PMN~J0948+0022. 

The optical part of the SED with simultaneous data from OM and REM has a somewhat irregular shape. While the statistical errors on the flux densities are low, there may be systematic calibration effects contributing to this behaviour. In particular, the optical instruments have been calibrated for stars rather than blazars and have not been cross-calibrated with each other.

\section{Discussion}
\label{discussion}

Below we discuss our results on \source in the context of the other $\gamma$-ray NLSy1s. After addressing the origin of the X-ray emission (section \ref{xdiscussion}) and the multi-wavelength properties (section \ref{mwdiscussion}), we conclude with a discussion about the nature of these sources (section \ref{nature}).

\subsection{Origin of the X-ray emission}
\label{xdiscussion}

\subsubsection{The population of $\gamma$-ray NLSy1s in X-rays}

\begin{figure}
\resizebox{80mm}{!}{\includegraphics{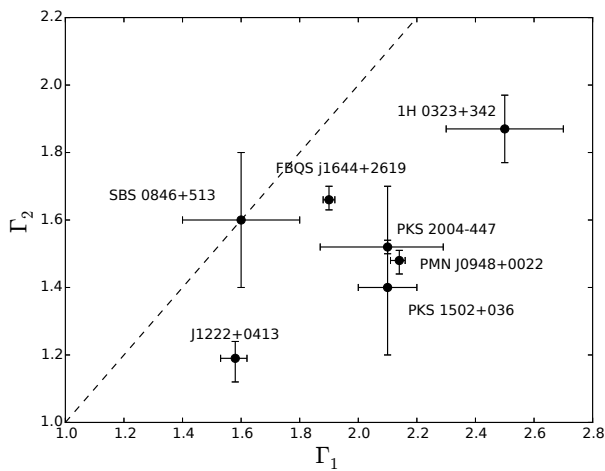}}
\caption{Photon indices below ($\Gamma_1$) and above ($\Gamma_2$) the break energy for the $\gamma$-ray NLSy1s. Data collected from \protect \cite{Gallo2006, deRosa2008,D'Ammando2013,D'Ammando2014, Foschini2015, Yao2015a}. See text for details. The case of equal photon indices below and above the break is indicted by the dashed line.}
\label{bknpow_all}
\end{figure}

Studies of the X-ray spectra of $\gamma$-ray NLSy1s clearly reveal some common trends.  In particular, the majority of sources have hard spectra above 2~keV, a soft excess at lower energies, and no evidence for intrinsic absorption.  A broken power law often provides a good fit, and we plot the photon indices below and above the break collected from the literature in Fig.~\ref{bknpow_all}. The plot includes seven out of the nine $\gamma$-ray NLSy1s detected to date. B3~1441+476  and NVSS~J124634+023808 are not included since they lack X-ray spectral information. The remaining seven sources are listed in  Section~\ref{introduction}. We note that the evidence for a soft excess is weak in PKS~2004-447, where there is a tentative detection in only one of the three \xmmn observations \citep{Gallo2006,Orienti2015,Kreikenbohm2016}. In addition, the X-ray spectrum of SBS~0846+513 is consistent with a single power law \citep{D'Ammando2013}. However, this source only has short \swift XRT observations, which are typically not sufficient to constrain more complex models (cf. Section~\ref{xrtresults}). For 1H~0323+342, we show in Fig.~\ref{bknpow_all} the photon indices obtained by fitting a broken power law to a spectrum simulated based on the double power-law model fitted to the {\it Suzaku} data in \cite{Yao2015a}. While this gives an indication of the spectral shape, a more complex spectrum was observed both in co-added {\it Swift}~XRT spectra, where hints of an Fe~line is seen \citep{Paliya2014}, and in a recent, deep \xmmn observation (\citealt{Kynoch2017}; D'Ammando et al.,  in prep).  For all the sources plotted in Fig.~\ref{bknpow_all}, the break energies are in the range 1.6--2.1~keV, with the exception of PKS~2004-447. In this source the break energy is $0.6$~keV in the first \xmmn observation \citep{Gallo2006} and $\sim$~2--3~keV in the later observations 
\citep{Orienti2015}, though the broken power law is not statistically significant in the latter case. 

As seen in Fig.~\ref{bknpow_all}, the photon indices above the break are in the range 1.2--1.9. This is significantly flatter than in radio-quiet NLSy1s ($\sim$~1.8--3.7, with a peak at $\sim 2.6$) and instead more similar to FSRQs ($\sim$~1.3--2.1 with a peak at $\sim 1.6$, see Fig. 2 of \citealt{Foschini2015}). This shows that the hard spectra of the $\gamma$-ray NLSy1s are likely dominated by Inverse Compton emission from the jets, just like FSRQs. This conclusion is also supported by the variability properties. While radio-quiet AGN commonly show a softer-when-brighter trend (\citealt{Sobolewska2009,Soldi2014}), the only  $\gamma$-ray NLSy1s for which this kind of variability has  been claimed is 1H~0323+342 (from \swift XRT observations, \citealt{Paliya2014}). However, the recent \xmmn observation of this source instead reveals a harder-when-brighter trend (D'Ammando et al.,  in prep). While the problem with background flares prevents us from placing strong constraints on the spectral variability of FBQS~J1644+2619, we find some weak evidence of harder-when-brighter variability (section \ref{xmmvar}). Spectral hardening with increasing X-ray flux has been reported in a number of $\gamma$-ray emitting blazars (e.g., \citealt{Gliozzi2006,Zhang2006,D'Ammando2011,Aleksic2014,Aleksic2015,Hayashida2015}), supporting the idea that it is indicative of emission form a jet.

\subsubsection{Origin of the soft excess}

A likely interpretation of the excess emission at low energies is that it has the same origin as the soft excess seen in regular Seyfert~1s. This component is often exceptionally strong in NLSy1s (e.g., \citealt{Ponti2010,Jin2013}), making it plausible that it would be detectable in the $\gamma$-ray emitting sources even though the jet emission is strong. The origin of the soft excess is debated, with different models often being able to fit the data equally well.  We find this to be the case also for FBQS~J1644+2619, where we obtain equally good fits with models that include reflection and Comptonisation in addition to the jet emission. It should be noted that the spectrum can in principle also be explained by a simple model consisting of power-law emission from a standard corona in addition to the jet. However, the very hard photon index inferred for the jet ($\Gamma=1.0^{+0.3}_{-0.4}$) and the resulting high predicted hard X-ray flux (in contention with the non-detection by {\it Swift}~BAT), makes this interpretation less likely.

In the reflection scenario, the power-law source is assumed to be a point source located on the rotational axis (modelled with {\sc relxilllp}), and from the fit we constrain the height of the source to be $h>11\ r_{\rm{g}}$. A more realistic model for the power-law source in our jet scenario would be an extended structure accelerated away from the disc, which was considered by  \cite{Dauser2013}. It was found that an extended structure is well approximated by a point source at an intermediate height, while acceleration means that the emission from the lowest part of the jet dominates the illumination of the disc. The constraint on the source height found in the fits to FBQS~J1644+2619 would thus correspond to the base of the jet. Since part of the emission from the point source directly reaches the observer, we are also implicitly assuming that the whole jet emits a spectrum with the same power-law slope. This is a simplification since the spectrum of the base of the jet, which illuminates the disc, may differ from the spectrum emitted further out, which dominates the direct emission seen by the observer. While this possibility could be accounted for by adding an additional power law to the model, the current observations do not allow us to constrain all the parameters of such a model.  From the fits we find that the disc has a low ionisation of~$(\xi)\ =1.6^{+0.3}_{-0.2}\ \ergcms$ and an inner edge $R_{\rm{in}} <42\ r_{\rm g}$. The latter constraint shows that the very innermost part of the disc does not contribute strongly to the reflection spectrum. This is a natural consequence of the large height of the illuminating source \citep{Dauser2013}, though an alternative interpretation would be that the standard disc is truncated further out than the ISCO (see further section \ref{nature} below). 

An important assumption of this model is that the main source of photons illuminating the disc is the base of the jet itself. This is motivated by the observational evidence that accretion disc coronae are compact (see \citealt{Uttley2014} and references therein) and the fact that compact coronae can naturally be associated with the base of a jet (in both X-ray binaries and AGN, e.g. \citealt{Markoff2005,King2017}). However, another possibility is that there is an additional, possibly extended, corona with a different power-law slope illuminating the disc.  The current observations do not allow us to constrain such a complex model, including contributions from a jet, corona and reflection. However, the simple double power-law model discussed above represents the case of corona+jet emission without any significant reflection. We also note that a possible scenario for jet launching is that an extended corona becomes collimated into a vertically extended structure \citep{Wilkins2015}, in which case the only primary source is the base of the jet, in line with our assumptions. 

The soft excess can also be modelled by Comptonisation of the thermal emission from the disc by a population of warm electrons (kT$\sim 0.3$~keV) with large optical depth ($\tau \sim 16$). These parameters are similar to what has been found in radio-quiet AGN where Comptonisation has been proposed as the most likely origin of the soft excess (e.g. \citealt{Petrucci2013,Mehdipour2015,Porquet2017}), as well the $\gamma$-ray NLSy1  PMN~J0948+0022 \citep{D'Ammando2014}. Even though we cannot discriminate between the reflection and Comptonisation scenarios in FBQS~J1644+2619,  both models imply the presence of an emission component similar to what is typically found in radio-quiet sources. A small number of FSRQs have previously been found to have contributions from both jet and Seyfert emission in their X-ray spectra \citep{Sambruna2006}. These sources were well described with broken power laws with parameters similar to FBQS~J1644+2619, but also exhibited weak Fe lines, lending support to the reflection interpretation. A scenario where different amounts of hard jet emission and softer Seyfert emission contribute to the X-ray spectra of $\gamma$-ray NLSy1s can likely explain some of the spread in photon indices in Fig.~\ref{bknpow_all}. 

Another possibility for the soft excess in $\gamma$-ray NLSy1s is that it is due to the jet itself. This may be the case if the tail of the synchrotron emission from the jet extends to the soft X-ray range, while the Inverse Compton component starts dominating at higher energies. However, modelling of the SEDs of these sources shows no indications that the synchrotron component should reach the X-ray band (e.g.~\citealt{Abdo2009b}). Another model that may explain the soft excess within a jet scenario involves Comptonisation by a shell of cold electrons moving along the jet \citep{Celotti2007}. However, such a feature is expected to be transient, in apparent contradiction with the fact that it is observed in the majority of the $\gamma$-ray NLSy1s. X-ray observations at different flux states is needed in order to further test different scenarios for the soft excess in these sources. If the soft excess originates from the accretion flow, its contribution to the X-ray spectrum should be stronger when the emission from the jet is lower. A low jet state should also make it possible to observe other reflection features, such as the Fe line. The small number of observations probing different flux states carried out to date are not sufficient to draw any clear conclusions regarding this. In 1~H0323+342 there are signs that the contribution from the jet is smaller at low X-ray fluxes \citep{Paliya2014}, while the weak soft excess in PKS~2004-447 was detected in the observation that had the highest X-ray flux \citep{Gallo2006,Orienti2015,Kreikenbohm2016}.

\subsection{Multiwavelength properties}
\label{mwdiscussion}

Modelling of the SEDs of  $\gamma$-ray NLSy1s has revealed them to be low-synchrotron peaked blazars, similar to FSRQs \citep{Abdo2009b, Foschini2012, D'Ammando2013, Yao2015a, D'Ammando2015a, Kynoch2017}. The high-energy emission in these sources is dominated by Inverse Compton scattering of photons external to the jet.  As discussed above, there is clear evidence that this component also contributes to the X-ray band below 10~keV (although the contribution in 1H~0323+342 seems negligible in some flux states, \citealt{Paliya2014}). We find that the SED of \source (Fig.~\ref{sed}) is consistent with this picture. The main difference in the SED compared to the other sources is the apparent $\gamma$-ray luminosity, which is low both in absolute terms (on average $1.6\times 10^{44}\ \rm erg\ s^{-1}$) and when compared to the luminosities in the radio--X-ray range. This indicates a lower Compton dominance in this source. Only 1H~0323+342 has a comparable $\gamma$-ray luminosity, while some of the other sources have average $\gamma$-ray luminosities larger by about three orders of magnitudes (\citealt{Abdo2009b}). 

In the radio band, \source has a flat spectrum, with a spectral index ($\alpha$, defined from $ S\propto \nu^{\alpha}$, where $S$ is the flux density) reported to be $\alpha = -0.07$  between 1.4 and 5~GHz \citep{Yuan2008} and $\alpha = +0.25$ between 1.7 and 8.4~GHz \citep{Doi2011}. This is similar to the other $\gamma$-ray NLSy1s \citep{Angelakis2015}. The Medicina observation presented here (Section \ref{medicina}) is the first observation of the source at 24~GHz. The observed flux density of $110\pm 30$~mJy is higher than the flux density of 62~mJy at 22 ~GHz reported by \cite{Doi2016} from a VLBI observation in 2014. However, these observations are not directly comparable due to the different angular resolutions involved.  As seen from the SED in Fig.~\ref{sed}, the Medicina observation is compatible with the archival radio data at lower and higher frequencies.  

From the monitoring with {\it Swift} and REM, we find \source to be variable in the X-ray, UV and optical bands on time-scales of days and months. The variability amplitude is larger in X-rays ($\sim 2.7$) than in the UV/optical ($\sim$~1.4--1.8). This is in line with the jet dominating the X-ray flux (it makes up $\sim 90$~per~cent of the flux according to the different models presented in Section \ref{xmmtav}), while the disc dominates the UV/optical emission and dilutes the variability. According to the accretion disc modelling presented by \cite{Calderone2013}, the ratio of jet to disc emission in the optical/UV range is $0.25$. We find some weak evidence for correlated variability between the X-ray and optical/UV fluxes, though the significance reaches $>2 \sigma$ only for the $b$-band. By comparison, clear evidence for correlated UV and X-ray variability on day time-scales was reported for the $\gamma$-ray NLSy1 1H~0323+342, which was interpreted as reprocessing of the X-ray emission by the accretion disc \citep{Yao2015a}.

\subsection{On the nature of $\gamma$-ray NLSy1s}
\label{nature}

The fact that powerful jets are commonly associated with massive elliptical galaxies has been interpreted as evidence that rapidly spinning black holes, resulting from major mergers, are needed for efficient jet formation \citep{Sikora2007}. On the other hand, NLSy1s usually have spiral host galaxies (e.g., \citealt{Deo2006}), which raises the question of whether such systems are able to produce the powerful jets seen in $\gamma$-ray NLSy1s. Only a small number of powerful radio sources have been associated with spiral galaxies to date (e.g. \citealt{Morganti2011,Singh2015}). Understanding the nature of the host galaxies of $\gamma$-ray NLSy1s is thus of great interest. So far, studies of the host galaxies have been performed only for the three most nearby sources, including FBQS~J1644+2619, which was studied in the infrared by \cite{Olgu2017} and \cite{D'Ammando2017}.  While \cite{Olgu2017} find that the host is likely a barred lenticular galaxy, \cite{D'Ammando2017} find that the data are consistent with an elliptical galaxy, using deeper observations. The other two host galaxies that have been studied are those of 1H0323+323, which exhibits an irregular ring morphology, possibly indicating a recent merger \citep{Leon2014}, and PKS~2004-447, for which  \cite{Kotilainen2016} reported a pseudo-bulge morphology. From this very small sample there is thus some indication that the hosts of $\gamma$-ray NLSy1s differ from the spiral galaxies typically associated with radio-quiet NLSy1s. 

A related issue is that of the black hole masses, with NLSy1s having lower masses ($\sim 10^6 - 10^7\ \Msun$, \citealt{Zhou2006}) compared to typical jetted AGN ($> 10^8\ \Msun$, \citealt{Chiaberge2011}). While it has been suggested that the black hole masses of NLSy1s may be underestimated as a result of flattened broad-line regions (BLR, \citealt{Decarli2008}) or effects of radiation pressure \citep{Marconi2008}, it is also the case that the X-ray variability supports the hypothesis of low masses in many cases (e.g., \citealt{Zhou2010}). For the $\gamma$-ray NLSy1s, the mass estimates are $\gtrsim 10^{7}\ \Msun$  \citep{Foschini2015, Yao2015b,Baldi2016,D'Ammando2017},  at the high end of the mass distribution for NLSy1s. It is notable that the inclination of the $\gamma$-ray emitting sources are known to be low, so if the virial mass estimates are affected by flattened BLRs (as seems to be the case for PKS~2004-447, \citealt{Baldi2016}), the difference with respect to the non-$\gamma$-ray emitting NLSy1s should be even greater. A further indication that the masses of the $\gamma$-ray NLSy1s may be underestimated comes from the jet scaling relations presented by \cite{Gardner2017}.  In the case of  FBQS~J1644+2619, virial mass estimates give $M_{\rm{BH}}=7.9\times10^6\ \rm{M_\odot}$ \citep{Yuan2008} and $M_{\rm{BH}}=1.4\times10^7\ \rm{M_\odot}$ \citep{Foschini2015}, while \cite{Calderone2013} finds  $M_{\rm{BH}}=2^{+6}_{-1} \times10^8\ \rm{M_\odot}$ from fitting a Shakura \& Sunyaev model for the accretion disc to optical and UV data, and \cite{D'Ammando2017} estimate $M_{\rm{BH}}=2.1 \pm 0.2 \times10^8\ \rm{M_\odot}$ from the bulge luminosity. 

Just like radio-quiet NLSys1, the radio-loud objects (including the $\gamma$-ray emitting ones) have been inferred to have high accretion rates \citep{Yuan2008}. This implies that jet formation may be occurring in a situation similar to the very high state of X-ray binaries \citep{Fender2004}. However, the estimates of the accretion rate for \source are relatively low, ranging between $0.007 - 0.2\ \times$ Eddington (\citealt{Calderone2013} and Section \ref{xmmtav}), with uncertainties arising from the black hole mass estimates and the contribution from the jet to the optical emission. The lowest of these values are in fact below the threshold where the inner part of the disc is usually assumed to transition into an Advection Dominated Accretion Flow (ADAF, \citealt{Narayan1995,Esin1997}). Such geometrically thick flows have in turn been linked to jet formation (e.g, \citealt{Livio1999}). The results of the X-ray spectral analysis (Sections \ref{xmmtav}, \ref{xdiscussion}) are also consistent with an ADAF in the very innermost part of the disc in FBQS~J1644+2619. Further studies of the population of $\gamma$-ray NLSy1s are needed to better constrain the properties of the accretion flows and the connection with jet formation in these sources. In summary, while there is growing evidence that the $\gamma$-ray NLSy1s may not be radically different from the blazar population as a whole, they do occupy the low end of the black hole mass distribution.

\section{Summary and conclusions}
\label{conclusions}

This paper presents a study of FBQS~J1644+2619, one of the newest members of the small class of $\gamma$-ray NLSy1s. We have analysed a deep \xmmn observation from March 2017, as well as quasi-simultaneous observations covering radio -- $\gamma$-rays obtained with the Medicina radio telescope, REM, {\it Swift} and {\it Fermi}~LAT. The main results can be summarized as follows:

\begin{itemize}

\item The \xmmn spectrum is characterised by a hard power law above ~2~keV and a soft excess at lower energies. The full 0.3--10~keV spectrum is well described by a broken power law with $\Gamma_1=1.90\pm 0.02$, $\Gamma_2=1.66^{+0.03}_{-0.04}$ and $E_{\rm break} =1.9^{+0.3}_{-0.2}$~keV. There is no evidence for intrinsic absorption and no detection of an Fe line. These properties are similar to the majority of the other $\gamma$-ray NLSy1s that have been studied in X-rays.  

\item The hard emission above $\sim$~2~keV is most likely dominated by inverse Compton emission from a jet, as in FSRQs. We also find weak evidence for harder-when-brighter variability, which is consistent with a strong contribution from a jet. 

\item A likely interpretation of the soft excess is that it is has a contribution from the underlying Seyfert emission. This contribution is equally well described by reflection from the base of the jet and by Comptonisation of disc emission in a warm, optically thick corona. The former model implies a relatively large height for the base of the jet ($h>11\ r_{\rm{g}}$) and no significant emission from the innermost part of the disc. 

\item  The monitoring observations in optical, UV and X-rays revealed variability on time-scales of days and months. The maximal variability amplitudes are $\sim$~1.4--1.8 in the optical/UV and $\sim 2.7$ in X-rays. 

\item The source was not detected in $\gamma$-rays a the time of the \xmmn observation. Considering a one-month interval, the 2$\sigma$ upper limit from {\it Fermi}~LAT in the 0.1--300~GeV energy range is 1.44$\times$10$^{-8}$ ph cm$^{-2}$ s$^{-1}$. This is consistent with previous findings that this source is at the low end of $\gamma$-ray fluxes observed in NLSy1s. 

\item Apart from the relatively low  $\gamma$-ray flux, the SED is similar to that observed in other $\gamma$-ray NLSy1s, confirming the blazar-like nature of the source. 
 
\end{itemize}

Finally, we note that \source differs from the vast majority of radio-quiet NLSy1s by having a relatively large black hole mass ($\sim 10^{7} - 2  \times10^8\ \rm{M_\odot}$), low accretion rate (0.007--0.2~ $\times$ Eddington) and a likely elliptical host galaxy. At the same time, the black hole mass is at the low end of the mass distribution for blazars. This, together with the presence of a soft X-ray excess, makes \source and the majority of the other $\gamma$-ray NLSy1s different from typical blazars.

\section*{Acknowledgements}

This work was supported by the Knut \& Alice Wallenberg foundation and the Swedish National Space Board. FD thanks S.
Covino for his help with the REM data reduction. The authors thank Eugenio Bottacini and Alberto Dominguez for helpful comments. 

The \textit{Fermi} LAT Collaboration acknowledges generous ongoing support
from a number of agencies and institutes that have supported both the
development and the operation of the LAT as well as scientific data analysis.
These include the National Aeronautics and Space Administration and the
Department of Energy in the United States, the Commissariat \`a l'Energie Atomique
and the Centre National de la Recherche Scientifique / Institut National de Physique
Nucl\'eaire et de Physique des Particules in France, the Agenzia Spaziale Italiana
and the Istituto Nazionale di Fisica Nucleare in Italy, the Ministry of Education,
Culture, Sports, Science and Technology (MEXT), High Energy Accelerator Research
Organization (KEK) and Japan Aerospace Exploration Agency (JAXA) in Japan, and
the K.~A.~Wallenberg Foundation, the Swedish Research Council and the
Swedish National Space Board in Sweden.
 
Additional support for science analysis during the operations phase is gratefully
acknowledged from the Istituto Nazionale di Astrofisica in Italy and the Centre
National d'\'Etudes Spatiales in France. This work performed in part under DOE
Contract DE-AC02-76SF00515.

Part of this work was based on observations with the Medicina telescope operated by INAF - Istituto di Radioastronomia.



\bsp	
\label{lastpage}
\end{document}